# The dual-path hypothesis for the emergence of anosognosia in Alzheimer's disease

Katia Andrade*[1,2], Thomas Guiyesse[1], Takfarinas Medani[3], Etienne Koechlin[4], Dimitrios Pantazis[5], Bruno Dubois[1,2]

Authors affiliations

[1] Institute of Memory and Alzheimer's Disease (IM2A), Department of Neurology, Assistance Publique-Hôpitaux de Paris (AP-HP), Sorbonne University, Pitié-Salpêtrière Hospital, 75013 Paris, France

[2] Frontlab, Paris Brain Institute (Institut du Cerveau, ICM), AP-HP, Pitié-Salpêtrière Hospital, 75013 Paris, France

[3] Signal & Image Processing Institute, University of Southern California, Los Angeles, CA 90089, USA

[4] École Normale Supérieure, Laboratoire de Neurosciences Cognitives et Computationnelles, 29 Rue d'Ulm, 75005 Paris, France

[5] McGovern Institute for Brain Research, Massachusetts Institute of Technology, Cambridge, MA, USA

*Correspondence to: Katia Andrade
Full address: Institut de la Mémoire et de la Maladie d'Alzheimer (IM2A)
Hôpital de la Pitié-Salpêtrière
47-83 boulevard de l'hôpital
75013 Paris
France
Phone number: +33 (0)1 42 16 75 17 47
E-mail: katia.santosandrade@gmail.com

**Abstract**

Although neurocognitive models have been proposed to explain anosognosia in Alzheimer's disease (AD), the neural cascade responsible for its origin in the human brain remains unknown. Here, we build on a mechanistic dual-path hypothesis that brings error-monitoring and emotional processing systems as key elements for self-awareness, with distinct impacts on the emergence of anosognosia in AD. Proceeding from the notion of anosognosia as a dimensional syndrome, ranging from the lack of concern about one's own deficits (i.e., anosodiaphoria) to the complete lack of awareness of deficits, our hypothesis states that (i) unawareness of deficits would result from a failure in the error-monitoring system, whereas (ii) anosodiaphoria would more likely result from an imbalance between emotional processing and error-monitoring systems. In the first case, a synaptic failure in the error-monitoring system, in which the cingulate cortex plays a major role, would have a negative impact on error (or deficits) awareness, preventing patients from becoming aware of their condition. In the second case, an impairment in the emotional processing system, in which the amygdala and orbitofrontal cortex play a major role, would prevent patients from monitoring the internal milieu for relevant errors (or deficits) and assigning appropriate value to them, thus biasing their impact on the error-monitoring system. Our hypothesis stems from two scientific premises. One comes from preliminary results in AD patients showing a synaptic failure in the error-monitoring system and decline of awareness at the time of diagnosis. Another comes from the somatic marker hypothesis, which proposes that emotional signals are critical to adaptive behavior. Further exploration will be of great interest to illuminate the foundations of self-awareness and improve our understanding of the underlying mechanisms of anosognosia in AD.

**Introduction**

Anosognosia – a term derived from the Greek: "a", absence; "nosos", disease; "gnosis", knowledge – was introduced more than a century ago by the French neurologist Joseph Babinski to describe the lack of awareness of a motor deficit resulting from right hemisphere damage (Langer and Levine, 2014). Since then, accumulating evidence has shown that anosognosia may affect any type of brain deficits or loss of function (McGlynn and Schacter, 1989; Mograbi and Morris, 2018) (Starkstein, 2014; Bastin *et al.*, 2021)(McGlynn and Schacter, 1989; Mograbi and Morris, 2018). For instance, it is now well established that unawareness for cognitive impairments is frequent from the early stages of Alzheimer's disease (AD) (Starkstein, 2014; Bastin *et al.*, 2021), with evidence that it may have a predictive value for worsening of cognition over the disease course (Starkstein *et al.*, 2006; Gerretsen *et al.*, 2017; Vannini, Amariglio, *et al.*, 2017). Moreover, anosognosia typically delays AD diagnosis and causes resistance to treatment and rehabilitation efforts (Clare, 2004; Cosentino *et al.*, 2011), increasing the burden of care (Turró-Garriga *et al.*, 2013).

Interestingly, sometimes patients appear aware of their deficits in explicit verbal reports, but show a lack of emotional concern about the deficits and act inappropriately given their condition. This behavior, called anosodiaphoria, has been related to anosognosia in a putative continuum that suggests a dimensional, quantitative, rather than a categorical, qualitative, syndrome, ranging from a lack of concern for the deficits (with implicit unawareness, probably depending on a pre-conscious mechanism) to a complete lack of awareness with explicit denial of the deficits (that is, an explicit unawareness). Like anosognosia, anosodiaphoria seems to increase with AD progression (Lindau and Bjork, 2014). Although underexplored, this hypothetical continuum suggests that emotional processing disturbances may be critical for anosognosia and, specially, for anosodiaphoria, possibly through the suppression of affective signals in response to one's failures that would normally trigger concern, worry, and appropriate adaptive behaviors. In line with this view, apathy – a motivational disorder characterized by loss of initiative and lack of emotional reactivity – has been associated with unawareness in AD, but unlike depression, greater apathy seems to correlate with higher levels of anosognosia (Mograbi and Morris, 2014; Starkstein, 2014; Azocar, Livingston and Huntley, 2021). Also, an association between unawareness of deficits and executive dysfunctions in AD has been consistently observed (Michon *et al.*, 1994; Amanzio *et al.*, 2011, 2013). More recently, by using distinct neuroimaging techniques, researchers also found that anosognosia for memory deficits in patients with prodromal AD may be the result of either functional metabolic changes or functional disconnection between brain regions supporting self-referential and memory processes (Perrotin *et al.*, 2015; Vannini, Hanseeuw, *et al.*, 2017). In addition, other researchers have established a link between structural lesions in the anterior cingulate cortex (ACC) and anosognosia in other non-memory cognitive domains since the early stages of AD (Valera-Bermejo *et al.*, 2020). Nevertheless, although the production of novel hypotheses explaining anosognosia have progressed in the last decades, there is still a lack of evidence for a mechanistic explanation of how anosognosia emerges in the brain.

The Cognitive Awareness Model (CAM) (Agnew and Morris, 1998), probably the most influential model of anosognosia in AD, is based on a modular framework that takes account of multiple levels in which (un)awareness phenomena could be generated, including sensory levels and differentiated levels within a hierarchy of memory consolidation processes. Specifically, by considering anosognosia as a failure of memory consolidation, the CAM predicts that several dissociations can be found in the relationship between awareness and memory, which would result in three main types of anosognosia: 1) Mnemonic anosognosia, reflecting a deficit in the consolidation of new information in the personal data base, which is at the origin of the metaphorical "*Petrified Self*" concept (Mograbi, Brown and Morris, 2009; Lenzoni, Morris and Mograbi, 2020); 2) Executive anosognosia*,* reflecting an impairment in the mechanism that normally allows for comparison between the actual performance and the stored past information; and 3) Primary anosognosia*,* reflecting a dysfunction within the metacognitive awareness system, which exists at the top of the CAM hierarchy. Later research integrated the role of emotional processes in the CAM (Morris and Mograbi, 2013), building upon earlier work (Rosen, 2011). For more information on this model, see also (Tagai *et al.*, 2020) and (Lenzoni, Morris and Mograbi, 2020).

Our hypothesis derives from a distinct perspective: the core of our rationale is that anosognosia in AD, rather than being modular or domain-specific, would emerge from a breakdown in the system responsible for error detection and awareness. These errors could be committed in the context of any type of deficits, depending on their level of impairment. Specifically, we predict that AD patients *with* anosognosia would have a critical synaptic failure in the error-monitoring system, while AD patients *without* anosognosia would have this system still intact, or at least able to compensate through a "more firing, less wiring" mechanism (Daselaar *et al.*, 2015). As such, anosognosia would result from a necessary, critical error-monitoring failure, affecting patients' awareness for any type of deficits, following their level of impairment, as a consequence of (1) *direct* or (2) *indirect* damage to that system, as detailed below.

**The dual-path hypothesis for the emergence of anosognosia in AD**

The *dual-path hypothesis* predicts that the lack of awareness, with explicit denial of deficits (i.e., anosognosia, or explicit unawareness; *first case*) would result from *direct* damage to the error-monitoring system; whereas the lack of concern of deficits (i.e., anosodiaphoria, or implicit unawareness; *second case*) would more likely result from a disturbance in the emotional processing system, with *indirect* impact on error-monitoring. These (1) *direct* and/or (2) *indirect paths* would be at the origin of a neural mechanistic cascade leading to a critical synaptic failure in the error-monitoring system, and would eventually result in the inability of subjects to perceive (explicitly and/or implicitly) their own errors or deficits in any domain (i.e., cognitive, behavioral, functional, etc.).

Specifically, our *mechanistic predictions* are the following:

In the *first case,* a failure in the error-monitoring system would have a *direct* impact on error (or deficit) awareness, thus preventing patients from becoming aware of their condition. Such a failure would reflect local damage to the cingulate cortex, with particular focus on its anterior (ACC) and posterior (PCC) parts, which are the brain generators of error-related potentials associated with error monitoring, namely the error-related negativity (ERN) for preconscious error detection, and the positivity error (Pe) for error awareness (O'Connell *et al.*, 2007) (**Figure 1A**). However, even in the case of a failure in the error-monitoring system, a preserved emotional processing system would assure some implicit awareness, with possible benefits on implicit learning and behavioral adjustment. In line with this view, recent evidence has found some implicit recognition of difficulties in AD patients despite their inability to explicitly estimate their own cognitive functioning (Geurten, Salmon and Bastin, 2021).

In the *second case,* an impairment in the emotional processing system, in which the amygdala (Amy) and orbitofrontal cortex (OFC) play a major role (Šimić *et al.*, 2021; Rolls *et al.*, 2023), would have an *indirect* impact on error-monitoring by rendering patients unable to detect relevant errors (or deficits) in the internal milieu, and to assign appropriate value to them. More precisely, such an impairment would bias the impact of deficits on the error-monitoring system, thus possibly affecting the generation of error-related potentials, and particularly the ERN, a preconscious error detection biomarker, whose amplitude appears also modulated by motivational and emotional factors (Hajcak, McDonald and Simons, 2004). Patients would therefore suffer from an implicit rather than explicit unawareness, being able to identify their own deficits though unable to understand their consequences and adapt their behavior. As such, local damage to the Amy and the OFC, but also structural and/or functional disconnections between these two regions, as well as between them and key brain regions of the error-monitoring system, could be at the origin of this implicit unawareness. In particular, these disconnections could result from (i) fiber integrity impairments in two main white matter tracts, namely the *uncinate fasciculus (Thiebaut de Schotten et al., 2012)*, relying the OFC (key brain region where somatic markers are generated from secondary emotions, particularly implicated in reward and affective value of stimuli (Šimić *et al.*, 2021)) to the Amy (key brain region where somatic markers are generated from primary emotions, particularly implicated in emotional arousal processes (Šimić *et al.*, 2021)); and the *cingulate bundle* (Bubb, Metzler-Baddeley and Aggleton, 2018), a complex brain network, containing not only long fibers linking the cingulate cortex (including the ACC and the PCC, implicated in error monitoring) with the frontal and temporal lobes, but also short association fibers connecting adjacent cortices; and/or (ii) a functional imbalance within and between two resting state brain networks, namely the *Salience Network* (SN) (Seeley *et al.*, 2007), anchored in the ACC, the Amy and other limbic structures, critically involved in the detection and response to relevant stimuli, and the *Default Mode Network* (DMN) (Raichle *et al.*, 2001), whose brain core is the PCC (Leech *et al.*, 2011), principally involved in self-referential processes (Jilka *et al.*, 2014) (**Figure 1B**). Importantly, the bases of impaired self- awareness and anosognosia have been closely linked to DMN functioning in AD (Antoine *et al.*, 2019; Mondragón, Maurits and De Deyn, 2019).

A *third prediction*, following a synergistic A plus B mechanism (**Figure 1A plus 1B**), would result in the most severe case of anosognosia, affecting both explicit and implicit awareness, with pronounced consequences, up to and including a full inability of patients to learn from their errors (or deficits) and to adapt their behavior.

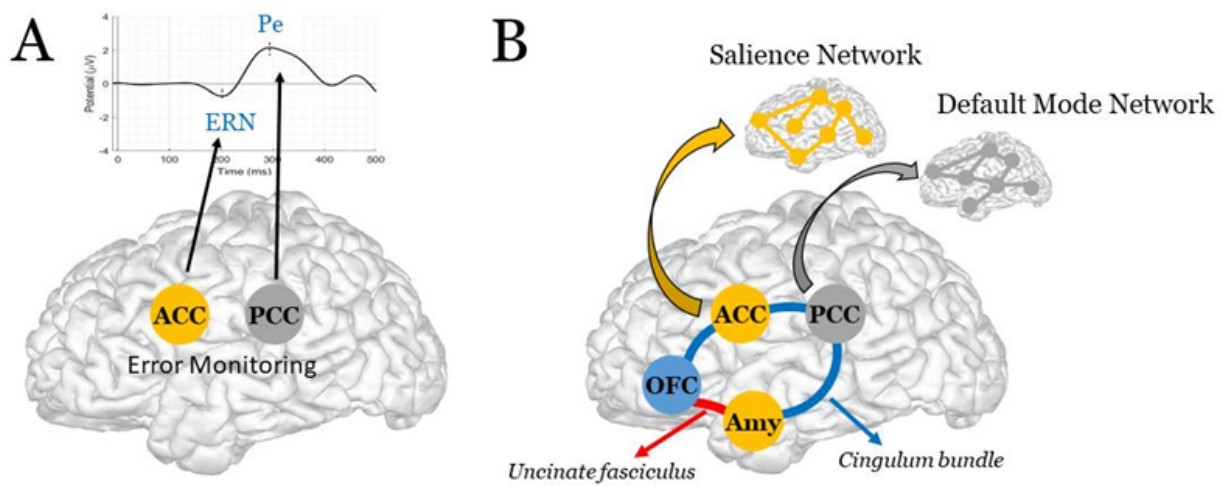

**Figure 1:** Schematic representation of the *Dual-path hypothesis for the emergence of anosognosia in AD*. **(A,** *Direct path***)** Synaptic failure in the error-monitoring system, due to direct damage to the ACC and/or the PCC, the brain generators of the ERN and the Pe (O'Connell *et al.*, 2007), with negative impact on explicit awareness. Of note, reduced metabolism has been reported in the PCC since the early stages of AD (Nestor *et al.*, 2003), with further evidence of greater hypometabolism in AD patients whose the onset of symptoms was before 65 years old (Rabinovici *et al.*, 2010). **(B,** *Indirect path***)** Structural and/or functional disconnections between the emotional processing and the error-monitoring systems, with negative impact on implicit awareness. The schematic diagram illustrates the cingulate fibers that travel along the anterior–posterior axis of the *cingulum bundle* (in blue) to reach various brain regions, including the medial temporal lobe, the Amy, and the OFC (Bubb, Metzler-Baddeley and Aggleton, 2018). The Amy and OFC are further interconnected via the *uncinate fasciculus* (in red) (Thiebaut de Schotten *et al.*, 2012). Of note, these two major white matter tracts have been shown to be impaired in the early stage of Alzheimer's disease (Kiuchi *et al.*, 2009). Further illustrated are the ACC and the Amy, which constitute key brain areas of the SN; and the PCC, which constitutes a key region of the DMN (see (Menon, 2011) for literature review on large-scale brain networks in psychopathology, highlighting potential parallels across neurological and psychiatric disorders).

**Scientific premises and evaluation of the hypothesis**

Our hypothesis stems on *two scientific premises*: the *one* comes from preliminary results showing a failure of the error-monitoring system in AD patients at the time of the diagnosis. Specifically, by studying error-related potentials during a word memory recognition task in two groups of cognitively normal amyloid-positive individuals at baseline, we have recently shown direct evidence of an error-monitoring failure, along with a cognitive awareness decline, only in subjects who progressed to AD during the five-year study period. In particular, we measured the ERN, mainly related to error detection, and the Pe, mainly related to error awareness, during a word memory recognition task in 51 amyloid positive individuals who had only subjective memory complaints at study entry. Of these, 15 individuals progressed to AD within the five-year study period (Prog group), and 36 remained cognitively normal (Stable group). We observed opposite longitudinal effects for the Prog and Stable groups. Specifically, we found a reduction of the Pe amplitude for the Prog group over time, in contrast with the increase of the Pe amplitude for the Stable group. Importantly, contrary to the Stable group, subjects who progressed to AD showed a cognitive awareness decline, with signs of anosognosia for their cognitive deficits at the moment of their AD diagnosis (Razafimahatratra *et al.*, 2022). Importantly, there is evidence that error-monitoring impairments in AD patients, and consequent unawareness of errors (including memory recognition deficits), are not merely a byproduct of their typical memory impairment (Dodson *et al.*, 2011).

The *second* premise comes from the "somatic marker hypothesis", proposed by Damasio (Damasio, 1996, no date). In fact, our rationale converges to some extent with that hypothesis, which states that one must "feel" the consequences of one's own actions, assigning them an affective value, in order to make the right decisions. In particular, Damasio proposes that decision-making (like self-awareness in the case of our *dual-path hypothesis*) requires the interplay between two specific brain systems: the executive and the emotional processing systems, in which the OFC (Bechara, Damasio and Damasio, 2000) and particularly the amygdala (Bechara, Damasio and Damasio, 2003) are necessary for triggering somatic states (Šimić *et al.*, 2021). Specifically, Damasio argues that autonomic reactions, such as electrodermal responses (EDR; see (Sequeira *et al.*, 2009), for literature review on electrical autonomic correlates of emotion) to stimuli, might prepare the subject to adapt attentionally and physically to changes in the environment (Damasio, 1996) Other researchers have also pointed to an important role for emotional dysregulation in producing unawareness, as errors may require an affective signature to motivate self-monitoring (Rosen, 2011). Supporting evidence exists for severe AD pathology in autonomic-related cortices, such as the OFC, which suggests that it could contribute to the emotional and autonomic dysregulations that often accompany this neurodegenerative disorder (Chu *et al.*, 1997).

In line with these premises, anosognosia of memory deficits has been associated with either hypoperfusion or hypometabolism in the PCC (Hanyu *et al.*, 2008; Perrotin *et al.*, 2015; Gerretsen *et al.*, 2017), the ACC (Hanyu *et al.*, 2008), and the OFC (Hanyu *et al.*, 2008; Perrotin *et al.*, 2015). Accordingly, additional evidence has shown reduced within- and between-network connectivity in the DMN in AD patients with anosognosia (Mondragón, Maurits and De Deyn, 2019, 2021), with an interesting association between hypometabolism in this network and an increased risk of progression to dementia in anosognosic patients (Therriault *et al.*, 2018).

Interestingly, evidence from another study found an association between memory monitoring and motor monitoring in AD patients, but observed that anosognosia for memory deficits was associated only with memory monitoring, not motor monitoring (Chapman *et al.*, 2018). The authors interpreted their results within a hierarchical model of awareness, suggesting that local self-monitoring processes were associated across different domains, but only contributed to overall levels of awareness in a domain-specific manner. We interpret their results in a different way. Our hypothesis predicts that anosognosia is not domain specific, but rather emerges when deficits are present in a given domain (thereby predisposing to error commission). In early stages of AD, with a typical amnestic presentation, a dysfunction in error-monitoring would largely concern memory deficits, rather than other less affected or unaffected functions, explaining why overall unawareness (or anosognosia for memory deficits as assessed through an offline, clinical interview method in that study) would be related to a failure in memory monitoring, but not to motor monitoring. Importantly, such failure in the error-monitoring system could also explain why anosognosia tends to worsen during the course of AD, following the level of impairment (and consequently the probability of making errors) in various cognitive, behavioral and functional domains beyond memory.

A new line of research is therefore needed to explore the *dual-path hypothesis for the emergence of anosognosia*. We aim at investigating this hypothesis based on within and between groups' comparisons, namely a group of AD patients, at distinct stages of the disease, presenting different levels of anosognosia, *versus* a group of healthy elderly controls, while performing a computer-based error-monitoring task. We will particularly focus our research on the study of erroneous responses and their potential correlations with the level of anosognosia and both central (namely, the ERN and the Pe) and peripheral (namely, the EDR) markers of error awareness and emotional arousal, measured simultaneously during the computer-based task. We will further investigate to what extent the level of anosognosia can be correlated with the amplitudes of these electrophysiological biomarkers, as well as with our hypothesis' main regions of interest (as illustrated in **Figure 1**: the PCC, the ACC, the OFC and the Amy, which have been associated with neural mechanisms of autonomic, affective, and cognitive integration (Bechara, Damasio and Damasio, 2000, 2003; Critchley, 2005)), through structural and functional neuroimaging methods.

**Consequences of the hypothesis and perspectives**

In sum, we hypothesize that anosognosia might emerge from a critical synaptic failure in the error-monitoring system, thus preventing patients from detecting (explicitly and/or implicitly) their own errors (or deficits). This failure could result from either (1) direct damage to the error-monitoring system (i.e., *Direct path*, **Figure 1A**) or from (2) the lack of emotional feedback on errors arriving to that system (i.e., *Indirect path*, **Figure 1B**), reflecting, in this case, local impairment in key structures of the emotional processing system and/or a disconnection between this system and the error-monitoring system.

Importantly, we have focused our hypothesis on AD as a pathological model, but our rationale implies that it can be applied to any other neurological condition in which anosognosia occurs. Indeed, anosognosia is prone to main deficits, according to their level of impairment, in several brain disorders. For example, anosognosia is mostly related to episodic memory deficits in the early stages of AD, whereas it can also affect other deficits following the severity of the disease (Leicht, Berwig and Gertz, 2010); as it is also mostly related to personality changes in the behavioral variant of frontotemporal dementia (Rankin *et al.*, 2005); or to hemiplegia, particularly in the acute phase after a right-hemisphere stroke (Orfei, Caltagirone and Spalletta, 2009); etc. Altogether, this strongly suggests that a common inability to monitor errors (or deficits), more than a specific consolidation impairment typical of AD, may be at the origin of this intriguing syndrome, and this regardless of the type of deficits (cognitive, behavioral, motor, etc.) or the neurological condition (neurodegenerative disorder, stroke, etc).

For instance, in the context of stroke, there is evidence of action-monitoring deficits in patients with anosognosia for hemiplegia (AHP) (Saj, Vocat, and Vuilleumier 2014; Vocat, Saj, and Vuilleumier 2013). Interestingly, these action-monitoring deficits seem to relate to monitoring deficits in distinct cognitive domains (Jenkinson *et al.*, 2009), thus supporting the existence of an error-monitoring impairment in stroke patients with AHP. Moreover, there is also evidence of an inability of these patients to monitor self-performed actions, while able to monitor others' actions or their own actions as if they were a third person (Fotopoulou *et al.*, 2009), which seems to indicate impairment in their self-referential systems. Of note, the DMN – and the PCC, as a central node of this network – play a key role in self-referential processes, with additional evidence indicating that damage to brain white matter tracts involved in these processes may foster the appearance of anosognosia in stroke patients (Pacella *et al.*, 2019).

Finally, elucidating the neural mechanistic cascade leading to anosognosia in AD may have two major clinical and scientific outcomes: *first,* contribute to a better understanding of the pathophysiology of this neurodegenerative disorder; *second,* refine current models of anosognosia with the goal of improving rehabilitation strategies allowing anosognosic patients to adhere to healthcare measures, which could maintain their autonomy for longer and reduce the burden of care. Importantly, as mentioned above, the rationale of our hypothesis extends beyond AD. Hence, to validate it, further research is required not only in AD patients but also in other neurological (and even psychiatric) populations, in which anosognosia has been frequently reported. Last, but not least, this line of research can shed new light on the theoretical foundations of human self-awareness.

## Author Contributions

KA: Conception of the hypothesis; Manuscript writing (first draft and final version). DP: Manuscript review and editing; Design of the "hypothesis figure", conceptualized with KA, with further contributions of TG and TM. All authors: Discussion of arguments in support of the hypothesis; Critical review of the final manuscript.

## Funding

This work was supported by a grant from the ANR (Agence Nationale de la Recherche): NOT_AWARE: ANR-17-CE37-0017-01 (to K.A.) and the National Institute of Aging of the National Institutes of Health under award number RF1AG074204 (to D.P.).

**Disclosure**

The authors declare that the present work was conducted in the absence of any commercial or financial relationships that could be construed as a potential conflict of interest.

**Consent statement/Ethical approval**

Not required.

**References**

Agnew, S.K. and Morris, R.G. (1998) 'The heterogeneity of anosognosia for memory impairment in Alzheimer's disease: A review of the literature and a proposed model', *Aging & mental health*, 2(1), pp. 7–19.

Amanzio, M. *et al.* (2011) 'Unawareness of deficits in Alzheimer's disease: role of the cingulate cortex', *Brain: a journal of neurology*, 134(Pt 4), pp. 1061–1076.

Amanzio, M. *et al.* (2013) 'Impaired Awareness of Deficits in Alzheimer's Disease: The Role of Everyday Executive Dysfunction', *Journal of the International Neuropsychological Society*, pp. 63–72. Available at: https://doi.org/10.1017/s1355617712000896.

Antoine, N. *et al.* (2019) 'Anosognosia and default mode subnetwork dysfunction in Alzheimer's disease', *Human brain mapping*, 40(18), pp. 5330–5340.

Azocar, I., Livingston, G. and Huntley, J. (2021) 'The Association Between Impaired Awareness and Depression, Anxiety, and Apathy in Mild to Moderate Alzheimer's Disease: A Systematic Review', *Frontiers in psychiatry / Frontiers Research Foundation*, 12, p. 633081.

Bastin, C. *et al.* (2021) 'Anosognosia in Mild Cognitive Impairment: Lack of Awareness of Memory Difficulties Characterizes Prodromal Alzheimer's Disease', *Frontiers in psychiatry / Frontiers Research Foundation*, 12, p. 631518.

Bechara, A., Damasio, H. and Damasio, A.R. (2000) 'Emotion, decision making and the orbitofrontal cortex', *Cerebral cortex* , 10(3), pp. 295–307.

Bechara, A., Damasio, H. and Damasio, A.R. (2003) 'Role of the amygdala in decision-making', *Annals of the New York Academy of Sciences*, 985(1), pp. 356–369.

Bubb, E.J., Metzler-Baddeley, C. and Aggleton, J.P. (2018) 'The cingulum bundle: Anatomy, function, and dysfunction', *Neuroscience and biobehavioral reviews*, 92, pp. 104–127.

Chapman, S. *et al.* (2018) 'Cross domain self-monitoring in anosognosia for memory loss in Alzheimer's disease', *Cortex; a journal devoted to the study of the nervous system and behavior*, 101, pp. 221–233.

Chu, C.C. *et al.* (1997) 'The autonomic-related cortex: pathology in Alzheimer's disease', *Cerebral cortex* , 7(1), pp. 86–95.

Clare, L. (2004) 'Awareness in early-stage Alzheimer's disease: A review of methods and evidence', *The British journal of clinical psychology / the British Psychological Society*, 43(2), pp. 177–196.

Cosentino, S. *et al.* (2011) 'Memory Awareness Influences Everyday Decision Making Capacity about Medication Management in Alzheimer's Disease', *International journal of Alzheimer's disease*, 2011, p. 483897.

Critchley, H.D. (2005) 'Neural mechanisms of autonomic, affective, and cognitive integration', *The Journal of comparative neurology*, 493(1), pp. 154–166.

Damasio, A. (no date) 'Descartes' Error: Emotion, Reason and the Human Brain, New York', *GP Putnam's Sons* [Preprint].

Damasio, A.R. (1996) 'The somatic marker hypothesis and the possible functions of the prefrontal cortex', *Philosophical transactions of the Royal Society of London. Series B, Biological sciences*, 351(1346), pp. 1413–1420.

Daselaar, S.M. *et al.* (2015) 'Less wiring, more firing: low-performing older adults compensate for impaired white matter with greater neural activity', *Cerebral cortex* , 25(4), pp. 983–990.

Dodson, C.S. *et al.* (2011) 'Alzheimer's disease and memory-monitoring impairment: Alzheimer's patients show a monitoring deficit that is greater than their accuracy deficit', *Neuropsychologia*, 49(9), pp. 2609–2618.

Fotopoulou, A. *et al.* (2009) 'Self-observation reinstates motor awareness in anosognosia for hemiplegia', *Neuropsychologia*, 47(5), pp. 1256–1260.

Gerretsen, P. *et al.* (2017) 'Anosognosia Is an Independent Predictor of Conversion From Mild Cognitive Impairment to Alzheimer's Disease and Is Associated With Reduced Brain Metabolism', *The Journal of clinical psychiatry*, 78(9), pp. e1187–e1196.

Geurten, M., Salmon, E. and Bastin, C. (2021) 'Impaired explicit self-awareness but preserved behavioral regulation in patients with Alzheimer disease', *Aging & mental health*, 25(1), pp. 142–148.

Hajcak, G., McDonald, N. and Simons, R.F. (2004) 'Error-related psychophysiology and negative affect', *Brain and cognition*, 56(2), pp. 189–197.

Hanyu, H. *et al.* (2008) 'Neuroanatomical correlates of unawareness of memory deficits in early Alzheimer's disease', *Dementia and geriatric cognitive disorders*, 25(4), pp. 347–353.

Jenkinson, P.M. *et al.* (2009) 'Reality monitoring in anosognosia for hemiplegia', *Consciousness and Cognition*, pp. 458–470. Available at: https://doi.org/10.1016/j.concog.2008.12.005.

Jilka, S.R. *et al.* (2014) 'Damage to the Salience Network and interactions with the Default Mode

Network', *The Journal of neuroscience: the official journal of the Society for Neuroscience*, 34(33), pp. 10798–10807.

Kiuchi, K. *et al.* (2009) 'Abnormalities of the uncinate fasciculus and posterior cingulate fasciculus in mild cognitive impairment and early Alzheimer's disease: A diffusion tensor tractography study', *Brain Research*, pp. 184–191. Available at: https://doi.org/10.1016/j.brainres.2009.06.052.

Langer, K.G. and Levine, D.N. (2014) 'Babinski, J.(1914). Contribution to the Study of the Mental Disorders in Hemiplegia of Organic Cerebral Origin (Anosognosia). Translated by KG Langer & DN Levine: Translated from the original Contribution à l'Étude des Troubles Mentaux dans l'Hémiplégie Organique Cérébrale (Anosognosie)', *Cortex; a journal devoted to the study of the nervous system and behavior* [Preprint]. Available at: https://psycnet.apa.org/record/2014-55242-003.

Leech, R. *et al.* (2011) 'Fractionating the default mode network: distinct contributions of the ventral and dorsal posterior cingulate cortex to cognitive control', *The Journal of neuroscience: the official journal of the Society for Neuroscience*, 31(9), pp. 3217–3224.

Leicht, H., Berwig, M. and Gertz, H.-J. (2010) 'Anosognosia in Alzheimer's disease: the role of impairment levels in assessment of insight across domains', *Journal of the International Neuropsychological Society: JINS*, 16(3), pp. 463–473.

Lenzoni, S., Morris, R.G. and Mograbi, D.C. (2020) 'The Petrified Self 10 Years After: Current Evidence for Mnemonic anosognosia', *Frontiers in psychology*, 11, p. 465.

Lindau, M. and Bjork, R. (2014) 'Anosognosia and anosodiaphoria in mild cognitive impairment and Alzheimer's disease', *Dementia and geriatric cognitive disorders extra*, 4(3), pp. 465–480.

McGlynn, S.M. and Schacter, D.L. (1989) 'Unawareness of deficits in neuropsychological syndromes', *Journal of clinical and experimental neuropsychology*, 11(2), pp. 143–205.

Menon, V. (2011) 'Large-scale brain networks and psychopathology: a unifying triple network model', *Trends in cognitive sciences*, 15(10), pp. 483–506.

Michon, A. *et al.* (1994) 'Relation of anosognosia to frontal lobe dysfunction in Alzheimer's disease', *Journal of neurology, neurosurgery, and psychiatry*, 57(7), pp. 805–809.

Mograbi, D.C., Brown, R.G. and Morris, R.G. (2009) 'Anosognosia in Alzheimer's disease – The petrified self', *Consciousness and Cognition*, pp. 989–1003. Available at: https://doi.org/10.1016/j.concog.2009.07.005.

Mograbi, D.C. and Morris, R.G. (2014) 'On the relation among mood, apathy, and anosognosia in Alzheimer's disease', *Journal of the International Neuropsychological Society: JINS*, 20(1), pp. 2–7.

Mograbi, D.C. and Morris, R.G. (2018) 'Anosognosia', *Cortex; a journal devoted to the study of the nervous system and behavior*, 103, pp. 385–386.

Mondragón, J.D., Maurits, N.M. and De Deyn, P.P. (2019) 'Functional Neural Correlates of Anosognosia in Mild Cognitive Impairment and Alzheimer's Disease: a Systematic Review', *Neuropsychology review*, 29(2), pp. 139–165.

Mondragón, J.D., Maurits, N.M. and De Deyn, P.P. (2021) 'Functional connectivity differences in Alzheimer's disease and amnestic mild cognitive impairment associated with AT(N) classification and anosognosia', *Neurobiology of Aging*, pp. 22–39. Available at: https://doi.org/10.1016/j.neurobiolaging.2020.12.021.

Morris, R.G. and Mograbi, D.C. (2013) 'Anosognosia, autobiographical memory and self knowledge in Alzheimer's disease', *Cortex; a journal devoted to the study of the nervous system and behavior*, 49(6), pp. 1553–1565.

Nestor, P.J. *et al.* (2003) 'Limbic hypometabolism in Alzheimer's disease and mild cognitive impairment', *Annals of neurology*, 54(3), pp. 343–351.

O'Connell, R.G. *et al.* (2007) 'The role of cingulate cortex in the detection of errors with and without awareness: a high-density electrical mapping study', *European Journal of Neuroscience*, pp. 2571–2579. Available at: https://doi.org/10.1111/j.1460-9568.2007.05477.x.

Orfei, M.D., Caltagirone, C. and Spalletta, G. (2009) 'The evaluation of anosognosia in stroke patients', *Cerebrovascular diseases* , 27(3), pp. 280–289.

Pacella, V. *et al.* (2019) 'Anosognosia for hemiplegia as a tripartite disconnection syndrome', *eLife*, 8. Available at: https://doi.org/10.7554/eLife.46075.

Perrotin, A. *et al.* (2015) 'Anosognosia in Alzheimer disease: Disconnection between memory and self-related brain networks', *Annals of neurology*, 78(3), pp. 477–486.

Rabinovici, G.D. *et al.* (2010) 'Increased metabolic vulnerability in early-onset Alzheimer's disease is not related to amyloid burden', *Brain: a journal of neurology*, 133(Pt 2), pp. 512–528.

Raichle, M.E. *et al.* (2001) 'A default mode of brain function', *Proceedings of the National Academy of Sciences of the United States of America*, 98(2), pp. 676–682.

Rankin, K.P. *et al.* (2005) 'Self awareness and personality change in dementia', *Journal of neurology, neurosurgery, and psychiatry*, 76(5), pp. 632–639.

Razafimahatratra, S. *et al.* (2022) 'Why are Alzheimer's disease patients unaware of their memory deficits?', *bioRxiv*. Available at: https://doi.org/10.1101/2022.02.01.478754.

Rolls, E.T. *et al.* (2023) 'Human amygdala compared to orbitofrontal cortex connectivity, and emotion', *Progress in neurobiology*, 220, p. 102385.

Rosen, H.J. (2011) 'Anosognosia in neurodegenerative disease', *Neurocase*, 17(3), pp. 231–241.

Seeley, W.W. *et al.* (2007) 'Dissociable intrinsic connectivity networks for salience processing and executive control', *The Journal of neuroscience: the official journal of the Society for Neuroscience*, 27(9), pp. 2349–2356.

Sequeira, H. *et al.* (2009) 'Electrical autonomic correlates of emotion', *International journal of psychophysiology: official journal of the International Organization of Psychophysiology*, 71(1), pp. 50–56.

Šimić, G. *et al.* (2021) 'Understanding Emotions: Origins and Roles of the Amygdala', *Biomolecules*, 11(6). Available at: https://doi.org/10.3390/biom11060823.

Starkstein, S.E. *et al.* (2006) 'A diagnostic formulation for anosognosia in Alzheimer's disease', *Journal of neurology, neurosurgery, and psychiatry*, 77(6), pp. 719–725.

Starkstein, S.E. (2014) 'Anosognosia in Alzheimer's disease: diagnosis, frequency, mechanism and clinical correlates', *Cortex; a journal devoted to the study of the nervous system and behavior*, 61, pp. 64–73.

Tagai, K. *et al.* (2020) 'Anosognosia in patients with Alzheimer's disease: current perspectives', *Psychogeriatrics: the official journal of the Japanese Psychogeriatric Society*, 20(3), pp. 345–352.

Therriault, J. *et al.* (2018) 'Anosognosia predicts default mode network hypometabolism and clinical progression to dementia', *Neurology*, 90(11), pp. e932–e939.

Thiebaut de Schotten, M. *et al.* (2012) 'Monkey to human comparative anatomy of the frontal lobe association tracts', *Cortex; a journal devoted to the study of the nervous system and behavior*, 48(1), pp. 82–96.

Turró-Garriga, O. *et al.* (2013) 'Burden associated with the presence of anosognosia in Alzheimer's disease', *International journal of geriatric psychiatry*, 28(3), pp. 291–297.

Valera-Bermejo, J.M. *et al.* (2020) 'Neuroanatomical and cognitive correlates of domain-specific anosognosia in early Alzheimer's disease', *Cortex; a journal devoted to the study of the nervous system and behavior*, 129, pp. 236–246.

Vannini, P., Hanseeuw, B., *et al.* (2017) 'Anosognosia for memory deficits in mild cognitive impairment: Insight into the neural mechanism using functional and molecular imaging', *NeuroImage: Clinical*, pp. 408–414. Available at: https://doi.org/10.1016/j.nicl.2017.05.020.

Vannini, P., Amariglio, R., *et al.* (2017) 'Memory self-awareness in the preclinical and prodromal stages of Alzheimer's disease', *Neuropsychologia*, 99, pp. 343–349.